\documentclass[showpacs,twocolumn,pra]{revtex4}
\usepackage{graphicx,amsfonts,amsmath,psfrag,bm}
\begin{document}  

\title{Polariton Analysis of a Four-Level Atom Strongly Coupled to a Cavity Mode}

\author{S. Rebi\'{c}}
\email[E-mail: ]{s.rebic@auckland.ac.nz} 
\author{A. S. Parkins}
\author{S. M. Tan}
\affiliation{Department of Physics, University of Auckland, Private Bag 92019, Auckland, New Zealand}

\begin{abstract}
We present a complete analytical solution for a single four-level atom strongly coupled to a cavity field mode and driven by external coherent laser fields. The four-level atomic system consists of a three-level subsystem in an EIT configuration, plus an additional atomic level; this system has been predicted to exhibit a photon blockade effect. The solution is presented in terms of polaritons. An effective Hamiltonian obtained by this procedure is analyzed from the viewpoint of an effective two-level system, and the dynamic Stark splitting of dressed states is discussed. The fluorescence spectrum of light exiting the cavity mode is analyzed and relevant transitions identified.
\end{abstract}


\maketitle

\section{Introduction}
\label{sec:intro}

The interaction of a single mode of the electromagnetic field with a single atom has long been at the forefront of interest within the quantum optics community. In this context, the Jaynes-Cummings model~\cite{Jaynes63} and its extensions have been the main focus of attention, for several reasons. It is the simplest possible model, involving a single two-level atom interacting with a quantized field mode, and therefore is in many cases exactly solvable. It has also proven to be experimentally realizable, thus allowing direct comparison between theory and experiment. This line of research has deepened immensely our understanding of fundamental quantum phenomena, and continues to do so.

Experimentally, the field of cavity quantum electrodynamics (CQED)~\cite{Berman94} has been shown to be very promising for further studies of fundamental quantum systems. Recent advances in mirror manufacturing techniques make it possible to build high-finesse microcavities in which the coupling strength of an atomic transition to a cavity field mode can be an order of magnitude larger than the decoherence rates of the system~\cite{Hood98}. Furthermore, four spectacular experiments have recently demonstrated that it is possible to trap a single atom within a microscopic cavity using either an independent atomic trap~\cite{Ye99,Guthohrlein01}, or the field mode itself, containing not more than one photon at a time~\cite{Hood00,Pinkse00}.

Within the framework of CQED, the regime of strong atom-field coupling is interesting for many reasons. For one, it enables the study of strongly coupled quantum systems; in particular, the `atom-cavity molecule'~\cite{Hood00}. Secondly, it is also a very promising candidate for the realization of strong optical nonlinearities~\cite{Dunstan98}. For example, an approximation to a $\chi^{(3)}$ (Kerr) nonlinear optical system can be achieved using either a single, strongly coupled two-level atom, or an ensemble of weakly coupled two-level atoms. However, the large atom-field detuning, which minimizes atomic spontaneous emission noise, also minimizes the strength of the nonlinearity. Using the Kerr-type nonlinearities produced by a single two-level atom in a cavity, conditional quantum dynamics have been demonstrated by Brune \textit{et al.}~\cite{Brune94} in the microwave regime, and by Turchette \textit{et al.}~\cite{Turchette95} in the optical regime.

The effect of electromagnetically induced transparency (EIT)~\cite{Harris97} has been used by Schmidt and Imamo\u{g}lu~\cite{Schmidt96} to devise a scheme involving four-level atoms which produces a large Kerr nonlinearity with virtually no noise. It has been shown by Imamo\u{g}lu \textit{et al.}~\cite{Imamoglu97} that if such a strong optical Kerr nonlinearity is implemented in a CQED setting, then it is possible to realize {\it photon blockade}, in which the atom-cavity system mimics an ideal two-level system, and effectively acts as a photon turnstile device for single photons. 

The proposal of Schmidt and Imamo\u{g}lu~\cite{Schmidt96} is very appealing in its use of EIT to substantially reduce decoherence. To utilize the advantages that a CQED environment offers, Rebi\'{c} \textit{et al.}~\cite{Rebic99} proposed a model in which a single four-level atom is trapped in a high-finesse microcavity. They showed that this system (which we call the EIT-Kerr system) can effect a near-ideal Kerr optical nonlinearity. In such a strongly coupled system, the composite excitations can be labeled as ``polaritons'', which are defined as mixtures of atom and cavity mode excitations. For weak to moderate driving, the EIT-Kerr system is well-approximated by a two-state system, corresponding to the two lowest lying polariton eigenstates. How this behaviour changes with the introduction of additional atoms was investigated by  Werner and Imamo\u{g}lu~\cite{Werner99} (see also the work of Greentree \textit{et al.}~\cite{Greentree00}).

Analyzing the single-atom EIT-Kerr system theoretically is not in general a straightforward task. In the bad cavity regime or the good cavity regime, approximate solutions are possible, based on the relative sizes of the atom-field coupling constant and the decay rates. In particular, it is possible to adiabatically eliminate either the cavity or the atomic degrees of freedom, respectively. In the strong coupling case, neither of these simplifications is possible. The `atom-field molecule' must be truly regarded as a fundamental entity, which exhibits features that cannot be explained in terms of individual properties of its constituents. The natural basis for analysis of such a system is the {\em polariton basis}. In this paper we perform a polariton analysis of the strongly coupled atom-cavity system. Although we concentrate on a particular atomic configuration, the underlying method is general and could be applied to any strongly coupled system.

Polariton analysis has been used extensively of late to study the dynamics of EIT systems~\cite{Fleischhauer00,Fleischhauer00b,Fleischhauer01}, but these analyses have concentrated on the semiclassical case of an atomic gas driven by laser light; in particular, on the dynamics of `slow' light. Juzeliunas and Carmichael~\cite{Juzeliunas01} have refined the analysis of the corresponding `slow polaritons', and showed that it is possible to reverse a stopped polariton by reversing the control beam. However, none of the treatments so far have dealt with the coupled atom-cavity system. 

In Section~\ref{sec:model}, we outline the bare model, i.e. the Hamiltonian written in terms of atomic and field operators, and explain how damping by reservoirs enters into the formulation. In Section~\ref{sec:dress} we diagonalise the interaction Hamiltonian exactly to find a set of basis states for subsequent analysis. In Section~\ref{sec:drivdamp}, the driving term and damping terms are expressed in terms of the new basis set, and the effective Hamiltonian in the polariton representation is found. In Section~\ref{sec:dynstark} we apply our results to obtain expressions for the dynamic Stark splitting and the spectrum of weak excitations in the effective two-level system. In Section~\ref{sec:fluorescence} we illustrate how to use the effective Hamiltonian to identify peaks and linewidths in the fluorescence spectrum for the light exiting the cavity mode. Finally, conclusions and outlook are presented in Section~\ref{sec:conclusion}.

\section{Bare Model}
\label{sec:model}

\begin{figure}[!t]
   \includegraphics[scale=0.75]{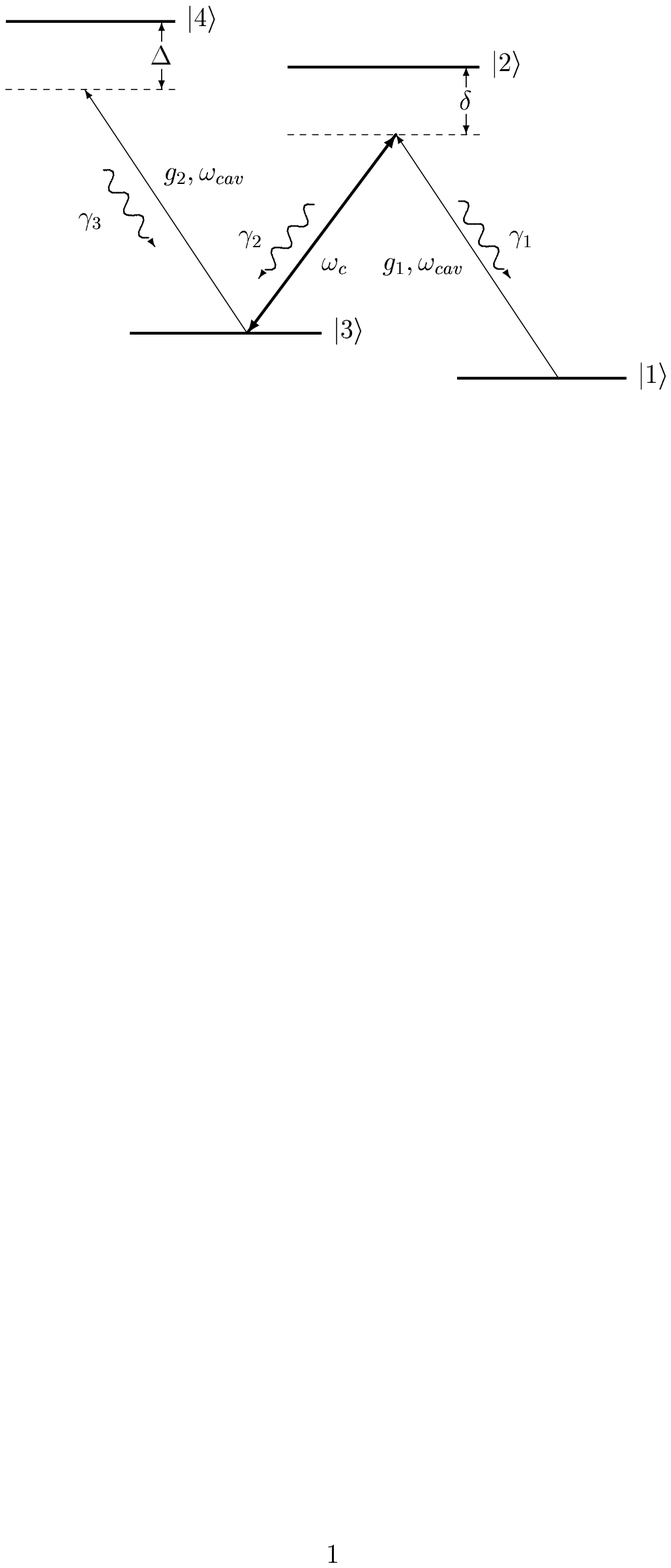}
   \caption{Atomic energy level scheme. The cavity mode couples to transitions $|1\rangle \rightarrow |2\rangle$ and $|3\rangle \rightarrow |4\rangle$, with respective coupling strengths $g_1$ and $g_2$. The transition $|2\rangle \leftrightarrow |3\rangle$ is coupled by a classical field of frequency $\omega_c$ and Rabi frequency $\Omega_c$. Spontaneous emission rates are denoted by $\gamma_j$. Detunings $\delta$ and $\Delta$ are defined as positive in the configuration shown.}
   \label{fig:atom}
\end{figure}

The atomic energy levels are shown in Fig.~\ref{fig:atom}. The atom is assumed to be coupled to a single cavity field mode and this cavity is driven through one of its mirrors by a coherent laser field. The interaction picture Hamiltonian describing the system in the rotating wave and electric dipole approximations is $\mathcal{H} = \mathcal{H}_0 + \mathcal{H}_d$, where
\begin{subequations}
 \label{eq:hamiltonian}
 \begin{eqnarray}
  \mathcal{H}_0 &=& \hbar\delta \, \sigma_{22} + \hbar\Delta \, \sigma_{44} + i\hbar g_1\, \bigl( a^\dagger \sigma_{12} - \sigma_{21} a \bigr) \nonumber \\ 
  &\ & + i\hbar \Omega_c\, \bigl( \sigma_{23} - \sigma_{32} \bigr) + i\hbar g_2\, \bigl( a^\dagger \sigma_{34} - \sigma_{43} a \bigr) , \label{eq:h0} \\
  \mathcal{H}_d &=& i\hbar \mathcal{E}_p \, \bigl( a - a^\dagger \bigr). \label{eq:hd} 
 \end{eqnarray}
\end{subequations}
Here, $\sigma_{ij}$ represent atomic raising and lowering operators (for $i \neq j$), and energy level population operators (for $i = j$); $a^\dagger \, (a)$ is the cavity field creation (annihilation) operator. Detunings $\delta$ and $\Delta$ are defined from the relevant atomic energy levels; $g_{1,2}$ are atom-field coupling constants for the respective transitions, and $\Omega_c$ is the coupling field Rabi frequency. The cavity driving field is introduced through the parameter ${\mathcal E}_p$, given by
\begin{equation}
 \label{eq:ep}
 \mathcal{E}_p = \sqrt{\frac{\mathcal{P} \kappa T^2}{4\hbar\omega_{cav}}}.
\end{equation}
In this expression, $T$ is the cavity mirror transmission coefficient, $\kappa$ is the cavity decay rate, and $\mathcal{P}$ is the power output of the driving laser. Damping due to cavity decay and spontaneous emission is discussed below.

Assume that the cavity mode subspace has been truncated at some finite size $N$. Together with the four atomic levels, these span a Hilbert space of dimension $4 \times N$. In the absence of driving (or in the limit where term~(\ref{eq:hd}) becomes negligible), Hamiltonian~(\ref{eq:h0}) takes a block-diagonal form, with $N$ blocks on the main diagonal. Each block represents a manifold of eigenstates associated with the appropriate term in the Fock expansion. The ground, first and second manifolds have been analyzed from the viewpoint of photon blockade in Refs.~\cite{Rebic99,Werner99,Greentree00}, where this truncation approach was found to be very useful. Addition of the driving term~(\ref{eq:hd}) significantly complicates the analysis. This term couples the different manifolds, and the Hamiltonian matrix loses its block-diagonal form. Therefore, it is not practical to perform a simple analytical diagonalization of the Hamiltonian~(\ref{eq:hamiltonian}), given the large size of the $4N$ by $4N$ matrix.

Dissipation can be added to the model by adding an anti-Hermitian term to the Hamiltonian~(\ref{eq:hamiltonian}). This term results from coupling to reservoir modes, and is obtained by tracing the system over these modes. In this approach we identify collapse operators, each of them corresponding to one decay channel~\cite{Carmichael93B}. In the EIT-Kerr case, there are the following four collapse operators
\begin{eqnarray}
  \label{eq:collapse}
 C_1 &=& \sqrt{\gamma_1}\, \sigma_{12}, \ \ C_2 = \sqrt{\gamma_2}\, \sigma_{32}, \nonumber \\
 C_3 &=& \sqrt{\gamma_3}\, \sigma_{34}, \ \ C_4 = \sqrt{\kappa}\, a\, ,
\end{eqnarray}
where $\gamma_k$ denote spontaneous emission rates into each of the decay channels, and $\kappa$ denotes the cavity intensity decay rate. The effective non-Hermitian Hamiltonian takes the form
\begin{equation}
 {\mathcal H}_{eff} = {\mathcal H} - i \hbar\sum_{k=1}^4 C_k^\dagger C_k \, , \label{eq:heff}
\end{equation}
${\mathcal H}$ being given by~(\ref{eq:hamiltonian}).

\section{Dressed States Analysis}
\label{sec:dress}

In this Section we solve the eigenvalue problem exactly for the Hamiltonian ${\mathcal H}_0$ given by Eq.~(\ref{eq:h0}), and obtain a basis for further calculations. In a strongly coupled system such as the one under analysis, dressed states~\cite{Cohen77} represent the natural basis for analysis, since the system under consideration should be viewed as an `atom-cavity molecule', rather than the mere sum of its constituent parts (atom + cavity mode in this case). We have already remarked in Section~\ref{sec:model} on the complexity of the problem of finding the exact (with coherent driving included) dressed states. Alsing {\em et al.}~\cite{Alsing92} succeeded in obtaining the exact solution for the case of a two-level atom when the driving field is resonant with the cavity mode. They recognized that the eigenstates can be expressed as a direct product of field and atomic states, where the field states are displaced squeezed states, thus simplifying the calculation. The driven EIT-Kerr system does not have a solution for the field states with similarly convenient properties, so the method of Ref.~\cite{Alsing92} can not be consistently applied. Instead we opt for an alternative approach which will be outlined in Section~\ref{sec:drivdamp}.

\subsection{Ground and First Manifold States}
\label{sec:gfman}

We use the notation $|${\em number of photons in cavity mode, atomic energy level} $\rangle$ to denote the bare states. The ground state is
\begin{equation}
  \label{eq:e00}
  |e_0^{(0)} \rangle = |0,1\rangle \, ,
\end{equation}
and has energy $E_0^{(0)} = 0$. Dressed state $j$ belonging to the manifold $n$ is denoted as $|e_j^{(n)} \rangle$.

There are three first-manifold states, one of them resonant with the cavity mode, the other two non-resonant
\begin{subequations}
  \label{eq:firstman}
  \begin{eqnarray}
    |e_0^{(1)} \rangle &=& \alpha_0^{(1)} |1,1\rangle + \mu_0^{(1)} |0,3\rangle \, , \label{eq:e01} \\
    |e_\pm^{(1)} \rangle &=& \alpha_\pm^{(1)} |1,1\rangle + \beta_\pm^{(1)}|0,2\rangle +  \mu_\pm^{(1)} |0,3\rangle \, , \label{eq:epm1}
  \end{eqnarray}
\end{subequations}
where the coefficients of the bare states are given by
\begin{subequations}
  \label{eq:firstcoeff}
  \begin{eqnarray}
    \alpha_0^{(1)} &=& \frac{1}{\sqrt{1+(g_1/\Omega_c)^2}}\, , \ \mu_0^{(1)} = \frac{g_1/\Omega_c}{\sqrt{1+(g_1/\Omega_c)^2}}\,  \label{eq:ce01}
  \end{eqnarray}
  \begin{eqnarray}
    \alpha_\pm^{(1)} &=& -\frac{g_1/\Omega_c}{\sqrt{1+(g_1/\Omega_c)^2+(\epsilon_\pm^{(1)}/\Omega_c)^2}}\, , \nonumber \\
    \beta_\pm^{(1)} &=& -\frac{i\epsilon_\pm^{(1)}/\Omega_c}{\sqrt{1+(g_1/\Omega_c)^2+(\epsilon_\pm^{(1)}/\Omega_c)^2}}\, , \ \\
    \mu_\pm^{(1)} &=& \frac{1}{\sqrt{1+(g_1/\Omega_c)^2+(\epsilon_\pm^{(1)}/\Omega_c)^2}}\, . \nonumber \label{eq:cepm1}
  \end{eqnarray}
\end{subequations}
The energies of these eigenstates are given by $E_j^{(1)} = \hbar(\omega_{cav} + \epsilon_j^{(1)})$, where 
\begin{subequations}
  \label{eq:en0pm}
  \begin{eqnarray}
    \epsilon_0^{(1)} &=& 0 \, , \\
    \epsilon_\pm^{(1)} &=& \frac{\delta}{2} \pm \sqrt{\biggl(\frac{\delta}{2} \biggr)^2 + \Omega_c^2 \left( 1 + \frac{g_1^2}{\Omega_c^2}\right)} \, .
\end{eqnarray}
\end{subequations}
Note that $\sum_{i=\pm,0} \epsilon_i^{(1)} = \delta$, reflecting the fact that the cavity mode is detuned from a one-photon excitation of the atom (see Fig.~\ref{fig:atom}).

\subsection{Second and Higher Manifold States}
\label{sec:secman}

The second and higher manifold states can be written in a generic form,
\begin{eqnarray}
  |e_k^{(n)}\rangle &=& \alpha_k^{(n)} |n,1\rangle + \beta_k^{(n)}|n-1,2\rangle \nonumber \\
  &\ &+ \mu_k^{(n)} |n-1,3\rangle  + \nu_k^{(n)} |n-2,4\rangle \, , \label{eq:ekn}
\end{eqnarray}
with $n \geq 2$ being the manifold label.

There are four states in each manifold, with energies $E_k^{(n)} = \hbar (n \omega_{cav} + \epsilon_k^{(n)})$. The coefficients of these states are
\begin{widetext}
\begin{subequations}
  \label{eq:secpluscoeff}
  \begin{eqnarray}
    \alpha_k^{(n)} &=& -i\, \frac{g_1g_2\sqrt{n(n-1)}}{\epsilon_k^{(n)}\Omega_c} \biggl[ 1-\frac{\epsilon_k^{(n)}(\epsilon_k^{(n)}-\Delta)}{g_2^2(n-1)} \biggr] \, \nu_k^{(n)} \, , \\
    \beta_k^{(n)} &=& \frac{g_2\sqrt{n-1}}{\Omega_c} \, \biggl[ 1-\frac{\epsilon_k^{(n)}(\epsilon_k^{(n)}-\Delta)}{g_2^2(n-1)} \biggr] \, \nu_k^{(n)} \, , \\
    \mu_k^{(n)} &=& -i\, \frac{\epsilon_k^{(n)}-\Delta}{g_2\sqrt{n-1}}\, \nu_k^{(n)} \, , \\
    \nu_k^{(n)} &=& \Biggl\{ 1 + \biggl(\frac{\epsilon_k^{(n)}-\Delta}{g_2\sqrt{n-1}} \biggr)^2 + \biggl( \frac{g_2\sqrt{n-1}}{\Omega_c} \biggr)^2 \biggl[ 1+n\biggl(\frac{g_1}{\epsilon_k^{(n)}} \biggr)^2\biggr] \biggl[ 1-\frac{\epsilon_k^{(n)}(\epsilon_k^{(n)}-\Delta)}{g_2^2(n-1)} \biggr]^2 \Biggr\}^{-1/2} \, .
  \end{eqnarray}
\end{subequations}
The exact energies of the four states within a given manifold are found to be, in increasing order,
\begin{subequations}
  \label{eq:epsilonkn}
  \begin{eqnarray}
    \epsilon_{1,2}^{(n)} &=& \frac{C}{4} - \frac{1}{2} \sqrt{\frac{C^2}{4}-\frac{2A}{3}+D} \mp \frac{1}{2} \sqrt{\frac{C^2}{4}-\frac{4A}{3}-D + \frac{2B+AC+C^3/4}{\sqrt{C^2/4-2A/3+D}}} \, , \\
    \epsilon_{3,4}^{(n)} &=& \frac{C}{4} + \frac{1}{2} \sqrt{\frac{C^2}{4}-\frac{2A}{3}+D} \mp \frac{1}{2} \sqrt{\frac{C^2}{4}-\frac{4A}{3}-D - \frac{2B+AC+C^3/4}{\sqrt{C^2/4-2A/3+D}}} \, ,
  \end{eqnarray}
\end{subequations}
where the following abbreviations have been used:
\begin{subequations}
  \label{eq:constants}
  \begin{eqnarray}
    A &=& \Delta\delta - g_1^2n - g_2^2(n-1) - \Omega_c^2 \, , \ C = \Delta + \delta \, , \\
    B &=& \Delta \bigl[ g_1^2n + \Omega_c^2 \bigr] + \delta \, g_2^2 (n-1) \, , \ G^2 = (g_1g_2)^2 \, n(n-1) \, , \\
    X_1 &=& 2A^3 + 9A(BC-G^2) + 27(B^2-C^2G^2) \, , \ X_2 = A^2 + 3BC +12G^2 \, , \\
    X &=& \sqrt[3]{X_1+\sqrt{X_1^2-4X_2^3}} \, , \ Y = X_2/X \, , \ D = \bigl( 2^{1/3}Y + 2^{-1/3}X \bigr)/3 \, .
  \end{eqnarray}
\end{subequations}
\end{widetext}
Note that for the $n$--th ($n \geq 2$) manifold, $\sum_{i=1}^4 \epsilon_i^{(n)} = \Delta +\delta$, which is the two-photon detuning of the atom from the cavity resonance. These equations are the exact eigenstates of Hamiltonian~(\ref{eq:h0}), and can be rewritten using the polariton operators as
\begin{eqnarray}
  \label{eq:h0pol}
  \mathcal{H}_0 &=& \hbar\epsilon_-^{(1)}\, p_-^{(1)\dagger}p_-^{(1)} + \hbar\epsilon_+^{(1)}\, p_+^{(1)\dagger}p_+^{(1)} \nonumber \\
  &\ & + \sum_{n=2}^\infty  \sum_{j=1}^4 \hbar\epsilon_j^{(n)}\, p_{kj}^{(n)\dagger}p_{kj}^{(n)} \, ,
\end{eqnarray}
where the polariton operators are defined as $p_{ij}^{(n)} = |e_i^{(n-1)}\rangle\langle e_j^{(n)}|$. Index $k$ in the second row of Eq.~(\ref{eq:h0pol}) is a dummy index, since $p_{kj}^{(n)\dagger}p_{kj}^{(n)} = |e_j^{(n)}\rangle\langle e_j^{(n)}|$. Polariton operators $p_{0,\pm}^{(1)} = |e_0^{(0)}\rangle\langle e_{0,\pm}^{(1)}|$ can be written in a relatively simple form. These expressions, as well as a short discussion on the statistical properties of polaritons, are given in Appendix~\ref{sec:app}.

\begin{figure}[!t]
   \includegraphics[scale=0.7]{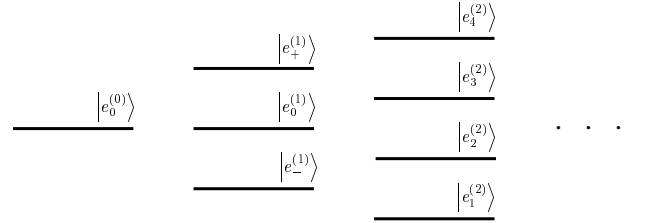}
   \caption{Schematic representation of the dressed states in a rotating system: ground, first and second manifold. The two states $|e_0^{(0)}\rangle$ and $|e_0^{(1)}\rangle$ with associated energies $\epsilon_0^{(0)} = 0$ and $\epsilon_0^{(1)} = 0$ are degenerate.}
   \label{fig:levels}
\end{figure}

The Hamiltonian~(\ref{eq:h0pol}) is written in a frame rotating at the cavity frequency $\omega_{cav}$. The energy level structure obtained has a form as shown in Fig.~\ref{fig:levels}: dressed states with the same detuning relative to the cavity resonance will be formally degenerate. There are only two degenerate eigenstates: ground state $|e_0^{(0)} \rangle$ and the first manifold state $|e_0^{(1)} \rangle$. It can be expected that driving this transition coherently yields dynamic Stark splitting. This indeed happens, and will be elaborated upon in Section~\ref{sec:dynstark}, but next we turn to the problem of how to include driving and damping into our polariton model.

\section{Driving and Damping Terms}
\label{sec:drivdamp}

The simplest way to include off-diagonal terms (as associated with driving and damping) into the effective Hamiltonian, written in the basis of states calculated in Section~\ref{sec:dress}, is to express the atomic and field operators in terms of operators describing the transitions between these states. Here again, we make a distinction between first and higher manifolds, a distinction imposed by the very different structure of these two groups of manifolds.

The first manifold dressed states can be expressed in terms of bare states as ${\mathbf b}_1 = {\mathbb M}_1{\mathbf d}_1$, where ${\mathbf b}_1 = \bigl( |1,\, 1\rangle,\, |0,\, 2\rangle\, , |0,\, 3\rangle \bigr)^{{\rm T}}$ and ${\mathbf d}_1 = \bigl( |e_-^{(1)}\rangle,\, |e_0^{(1)}\rangle,\, |e_+^{(1)}\rangle \bigr)^{{\rm T}}$. The transformation matrix is 
\begin{eqnarray}
  \label{eq:firstmatrix}
    {\mathbb M}_1 &=& \begin{pmatrix} -\frac{g_1/\Omega_c}{N_-} & \frac{1}{N_0} & -\frac{g_1/\Omega_c}{N_+} \\ i\frac{\epsilon_-^{(1)}/\Omega_c}{N_-} & 0 & i\frac{\epsilon_+^{(1)}/\Omega_c}{N_+} \\ \frac{1}{N_-} & \frac{g_1/\Omega_c}{N_0} & \frac{1}{N_+} \end{pmatrix} \, ,
\end{eqnarray}
where
\begin{subequations}
  \label{eq:ns}
\begin{eqnarray}
N_0 &=& \sqrt{1+\left( g_1/\Omega_c\right)^2}\, , \\
N_\pm &=& \sqrt{1+\left( \epsilon_\pm/\Omega_c \right)^2 + \left( g_1/\Omega_c\right)^2}\, .
\end{eqnarray}
\end{subequations}
Similarly, we can write for the higher manifolds ${\mathbf b}_n = {\mathbb M}_n{\mathbf d}_n$, $n \geq 2$, where ${\mathbf b}_n = \bigl( |n,\, 1\rangle,\, |n-1,\, 2\rangle\, , |n-1,\, 3\rangle\, , |n-2,\, 4\rangle \bigr)^{{\rm T}}$ and ${\mathbf d}_n = \bigl( |e_1^{(n)}\rangle,\, |e_2^{(n)}\rangle,\, |e_3^{(n)}\rangle,\, |e_4^{(n)}\rangle \bigr)^{{\rm T}}$, and
\begin{eqnarray}
  \label{eq:nthmatrix}
    {\mathbb M}_n &=& \begin{pmatrix}
      \alpha_1^{(n)*} & \alpha_2^{(n)*} & \alpha_3^{(n)*} & \alpha_4^{(n)*} \\
      \beta_1^{(n)*} & \beta_2^{(n)*} & \beta_3^{(n)*} & \beta_4^{(n)*} \\
      \mu_1^{(n)*} & \mu_2^{(n)*} & \mu_3^{(n)*} & \mu_4^{(n)*} \\
      \nu_1^{(n)*} & \nu_2^{(n)*} & \nu_3^{(n)*} & \nu_4^{(n)*} \end{pmatrix} \, ,
\end{eqnarray}
with the coefficients given by Eqs.~(\ref{eq:secpluscoeff}). These expressions provide all of the information needed for the subsequent calculations.

\subsection{External Driving}
\label{sec:extdrive}

Strongly coupled systems are very sensitive to the number of photons. In fact, the most interesting regimes include one or a few photons. In the system under investigation, the effect of photon blockade occurs when the dynamics is limited to the exchange of excitation between the ground state and the first manifold. It is therefore natural to express the field annihilation operator $a$ in terms of the transitions it produces between two adjacent manifolds. In general, deexcitation from manifold $n$ to manifold $n-1$ occurs via the operator $a^{(n)}$, expressed in terms of the bare states as
\begin{subequations}
  \label{eq:an}
\begin{eqnarray}
  &\, &a^{(1)} = |0,\, 1\rangle\langle 1,\, 1| \, \\
  &\, &a^{(2)} = \sqrt{2} |1,\, 1\rangle\langle 2,\, 1| + |0,\, 2\rangle\langle 1,\, 2| + |0,\, 3\rangle\langle 1,\, 3| \, \\
  &\, &a^{(n)} = \sqrt{n} |n-1,\, 1\rangle\langle n,\, 1| + \sqrt{n-1} \left( |n-2,\, 2\rangle\langle n-1,\, 2| \right. \nonumber \\
    &\ & \left. + |n-2,\, 3\rangle\langle n-1,\, 3| \right) + \sqrt{n-2} |n-3,\, 4\rangle\langle n-2,\, 4| \, ,
\end{eqnarray}
\end{subequations}
and the full annihilation operator would then be given by a sum over all manifolds, $a = \sum_{n=1}^\infty a^{(n)}$. For the transition from the first excited state to the ground state we obtain
\begin{subequations}
\label{eq:rabifirst}
\begin{eqnarray}
  \label{eq:a1}
  {\mathcal E}_p a^{(1)} = \Omega_-^{(1,0)} p_-^{(1)} + \Omega_0^{(1,0)} p_0^{(1)} + \Omega_+^{(1,0)} p_+^{(1)},
\end{eqnarray}
with the polariton operators defined by $|e_0^{(0)}\rangle = p_j^{(1)} |e_j^{(1)}\rangle,\ j = 0,\pm$. The effective Rabi frequencies $\Omega_j^{(0,1)}$ can be calculated from the matrix~(\ref{eq:firstmatrix}) as
\begin{eqnarray}
  \label{eq:omega01}
  \Omega_0^{(1,0)} &=& \frac{{\mathcal E}_p}{\sqrt{1+\bigl( g_1/\Omega_c\bigr)^2}} \label{eq:omega0} \, , \\
  \Omega_\pm^{(1,0)} &=& -\frac{{\mathcal E}_pg_1}{\sqrt{g_1^2+\Omega_c^2+(\epsilon_\pm^{(1)})^2}} \label{eq:omegapm} \, .
\end{eqnarray}
\end{subequations}
The three terms in the expansion~(\ref{eq:a1}) correspond to the three transitions between the ground state and the three states excited by a single photon. Each transition has an associated effective Rabi frequency $\Omega_j^{(1,0)}$. Note that the negative sign in~(\ref{eq:omegapm}) means the driving of the off-resonant states is out of phase with the driving of the resonant state (see Fig.~\ref{fig:manifolds} $(a)$).

There are twelve possible transitions between the first and the second manifolds, driven with the effective Rabi frequencies $\Omega_{ij}^{(2,1)}$, where
\begin{equation}
  \label{eq:omega12}
\Omega_{ij}^{(2,1)} = {\mathcal E}_p \left[ \sqrt{2}\, \alpha_i^{(1)*}\alpha_j^{(2)} + \beta_i^{(1)*}\beta_j^{(2)} + \mu_i^{(1)*}\mu_j^{(2)} \right] \, ,
\end{equation}
with $i = 0,\, \pm$; $j = 1,\ldots,4$ and it follows from the Eq.~(\ref{eq:e01}) that $\beta_0^{(1)} \equiv 0$ (see Fig.~\ref{fig:manifolds} $(b)$). Since the second and subsequent manifolds have four states each, there are sixteen transitions between the adjacent manifolds, with effective Rabi frequencies of driving
\begin{eqnarray}
  \label{eq:omegann}
\Omega_{ij}^{(n,n-1)} &=& {\mathcal E}_p \left[ \sqrt{n}\, \alpha_i^{(n-1)*}\alpha_j^{(n)} + \sqrt{n-1}\, \left( \beta_i^{(n-1)*}\beta_j^{(n)} \right. \right. \nonumber \\
    &\ &\left. \left. + \mu_i^{(n-1)*}\mu_j^{(n)} \right) + \sqrt{n-2}\, \nu_i^{(n-1)*}\nu_j^{(n)} \right] \, ,
\end{eqnarray}
with $n>2$ and $i,\, j = 1,\ldots,4$ (see Fig.~\ref{fig:manifolds} $(c)$). The coefficients in this expression are given in Eqs~(\ref{eq:secpluscoeff}).

Note that the Rabi frequencies $\Omega_{ij}^{(n,n-1)}$ can also be obtained from $i\hbar\Omega_{ij}^{(n,n-1)} = \langle e_i^{(n-1)}|{\mathcal H}_d|e_j^{(n)}\rangle$, with ${\mathcal H}_d$ given by Eq.~(\ref{eq:hd}). However, the expansion of the operator $a$ in terms of contributions to different transitions, Eq.~(\ref{eq:an}), offers a clearer physical picture of the processes involved in the dynamics. The driving Hamiltonian can therefore be written in terms of the polariton operators as
\begin{eqnarray}
  \label{eq:hdrivepol}
  {\mathcal H}_d &=& i\hbar \mathcal{E}_p \, \left( a - a^\dagger \right) \nonumber \\
  &=& i\hbar\sum_{i=\pm,0} \Omega_i^{(1,0)} \left( p_i^{(1)} - p_i^{(1)\dagger} \right) \nonumber \\
  &\ & +i\hbar\sum_{i=\pm,0} \sum_{j=1}^4 \Omega_{ij}^{(2,1)} \left( p_{ij}^{(2)} - p_{ij}^{(2)\dagger} \right) \nonumber \\
  &\ & + i\hbar\sum_{n=3}^\infty \sum_{i,j=1}^4 \Omega_{ij}^{(n,n-1)} \left( p_{ij}^{(n)} - p_{ij}^{(n)\dagger} \right) \, .
\end{eqnarray}

\begin{widetext}
Expression~(\ref{eq:hdrivepol}) is the expansion of the driving Hamiltonian in terms of the transitions that are permitted to occur between any two dressed states, as shown in Fig.~\ref{fig:manifolds}. The obvious advantage of this expansion over the original form of driving is in the strong coupling/low photon number regime. In this regime, the expansion~(\ref{eq:hdrivepol}) can be truncated at the order justified by the problem, while still retaining all (but not more!) of the relevant contributions from the external coherent driving. We will illustrate this assertion in Section~\ref{sec:dynstark}.
\begin{figure}[!t]
  \begin{center}
   \includegraphics[scale=1.2]{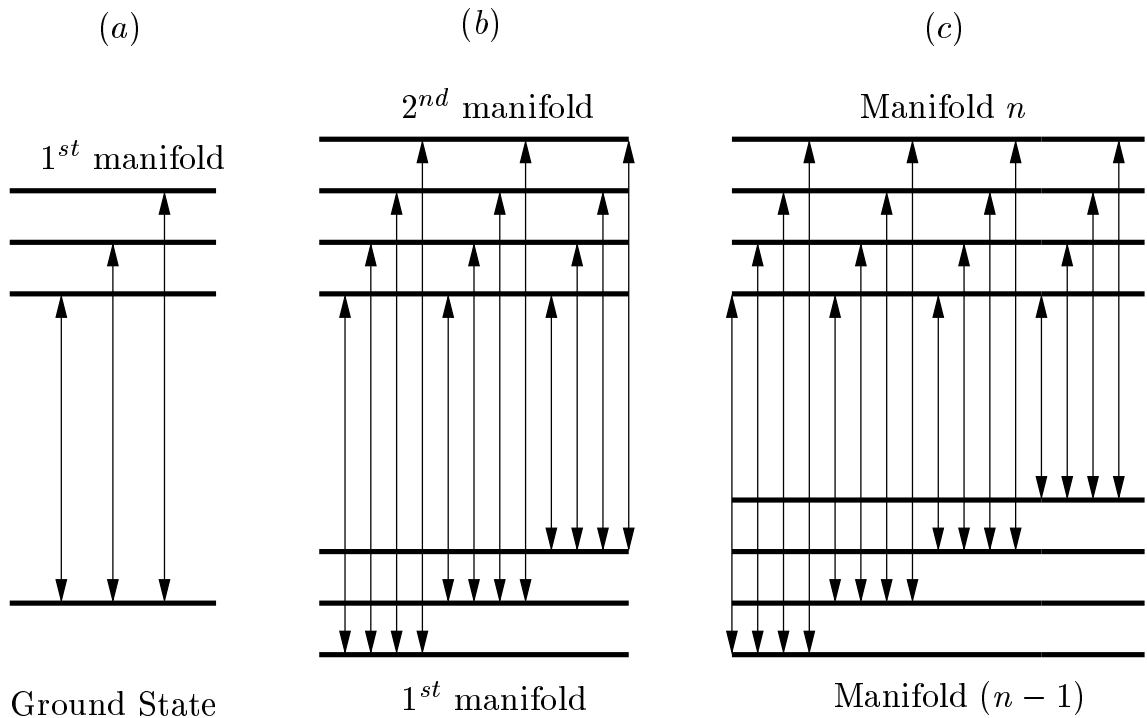}
   \end{center}
   \caption{Transition between the polaritons in the adjacent manifolds. The cavity resonance is located at the center of each  manifold. Figures represent: $(a)$ Transitions between the ground state and first manifold states; $(b)$ Transitions between first and second manifolds states; $(c)$ Transitions between polaritons in manifolds $(n-1)$ and $n$ for $n \geq 3$.}
   \label{fig:manifolds}
\end{figure}
\end{widetext}
\vspace{1cm}

\subsection{Damping by Reservoir Modes}

The remaining part of the dynamics to be expressed in the polariton representation is damping by the reservoir modes. In Section~\ref{sec:model} it was explained how the damping enters into the effective Hamiltonian; in particular after a trace has been performed over the reservoir variables. The resulting Hamiltonian operator is anti-Hermitian and has the form
\begin{equation}
  \label{eq:hres}
  {\mathcal H}_{res} = -i\hbar\kappa a^\dagger a -i\hbar(\gamma_1+\gamma_2) \sigma_{22} -i\hbar\gamma_3 \sigma_{44} \, .
\end{equation}
We follow the reasoning of the previous Section and expand the relevant field and atomic operators in terms of the contributions from the individual manifolds:
\begin{subequations}
\label{eq:dampops}
\begin{eqnarray}
  a^\dagger a &=& \sum_{n=1}^\infty \left( a^{(n)\dagger} a^{(n)} \right) \\
  &=& |1,\, 1\rangle\langle 1,\, 1| \nonumber \\
  &+& \sum_{n=2}^\infty \left[ n \, |n,\, 1\rangle\langle n,\, 1| + (n-1)\, \left( |n-1,\, 2\rangle\langle n-1,\, 2| \right. \right. \nonumber \\
    &+& \left. \left. |n-1,\, 3\rangle\langle n-1,\, 3| \right) + (n-2) |n-2,\, 4\rangle\langle n-2,\, 4| \right] \, , \nonumber 
\end{eqnarray}
\begin{eqnarray}
    \sigma_{22} &=& \sum_{n=1}^\infty \sigma_{22}^{(n)} = \sum_{n=1}^\infty |n-1,\, 2\rangle\langle n-1,\, 2| \, , \\
    \sigma_{44} &=& \sum_{n=2}^\infty \sigma_{44}^{(n)} = \sum_{n=2}^\infty |n-2,\, 4\rangle\langle n-2,\, 4| \, .
\end{eqnarray}
\end{subequations}
The operator ${\mathcal H}_{res}$ clearly takes a block-diagonal form in the dressed state representation, as the operator expansion~(\ref{eq:dampops}) includes terms containing every possible dressed level within a given manifold. The diagonal terms correspond to damping of the dressed states due to their decay straight into the reservoir. Off-diagonal terms couple two different dressed levels in a given manifold. This coupling arises due to couplings of both levels to the same reservoir level. If each of these diagonal blocks is again diagonalized, in the presence of damping we get shifts appearing on each level. The complex eigenvalues add their real part to the energy shift and the imaginary part becomes the damping rate. Energies and damping rates calculated in this manner will coincide with the experimentally observed ones (in the absence of driving). It was pointed out by Harris~\cite{Harris89} and Imamo\u{g}lu~\cite{Imamoglu89a} that these cross terms can be essential in creating destructive interference between the transition amplitudes of the appropriate transitions (see also Li and Xiao~\cite{Li95}).

The contribution of the off-diagonal terms to the eigenenergies and damping rates (diagonal elements of ${\mathcal H}_{res}$) is very small, and we will ignore their contribution to the eigenvalues in the rest of this paper for simplicity, though we leave them in a general expression for the damping Hamiltonian.

We write the damping Hamiltonian in a form that emphasizes the diagonal and off-diagonal contributions,
\begin{eqnarray}
  \label{eq:hrespol}
  {\mathcal H}_{res} &=&-i\hbar \sum_{i=\pm,0} \Gamma_i^{(1)} p^{(1)^\dagger}_{i} p^{(1)}_{i} \nonumber \\
  &\ & -i\hbar\sum_{j \neq k = \pm,0} \Gamma_{jk}^{(1)} p^{(1)^\dagger}_{j} p^{(1)}_{k} \nonumber \\
  &\ & -i\hbar\sum_{n=2}^\infty \sum_{j=1}^4 \Gamma_{jj}^{(n)} \, p^{(n)^\dagger}_{ij} p^{(n)}_{ij} \nonumber \\
  &\ & -i\hbar\sum_{n=2}^\infty \sum_{j \neq k} \Gamma_{jk}^{(n)} \, p^{(n)^\dagger}_{ij} p^{(n)}_{ik} \, .
\end{eqnarray}
where
\begin{subequations}
\begin{eqnarray}
  \label{eq:damprates}
  &\, &\Gamma_{jk}^{(n)} = n\kappa \, \alpha_j^{(n)*}\alpha_k^{(n)}  + \left[ (n-1)\kappa + \gamma_1 + \gamma_2 \right] \beta_j^{(n)*} \beta_k^{(n)} \nonumber \\
  &\ &+ (n-1)\kappa\, \mu_j^{(n)*} \mu_k^{(n)}  + \left[ (n-2)\kappa + \gamma_3 \right] \nu_j^{(n)*}\nu_k^{(n)} \, ,
\end{eqnarray}
\end{subequations}
For each manifold $(n)$, matrix $\Gamma_{jk}^{(n)}$ is a positive definite matrix, so we can write~\cite{Zhou97,Akram01}
\begin{subequations}
\begin{eqnarray}
  \label{eq:costheta}
  \Gamma_{jk}^{(n)} &=& \cos{\theta_{jk}}\, \sqrt{\Gamma_{jj}^{(n)}\Gamma_{kk}^{(n)}} \, , \\
  \cos{\theta_{jk}} &=& \frac{\bm{\mu}_j\cdot \bm{\mu}_k}{|\bm{\mu}_j||\bm{\mu}_k|} \, ,
\end{eqnarray}
\end{subequations}
where $\bm{\mu}_{j,k}$ can be thought of as the effective dipole moments of the transitions contributing the off-diagonal term. Furthermore, we note that the diagonal matrix elements belonging to the first manifold can be written in a simple closed form as
\begin{subequations}
  \label{eq:firstdamping}
\begin{eqnarray}
  \Gamma_0^{(1)} &=& \frac{\kappa}{1+\left( g_1/\Omega_c \right)^2} \label{eq:gamma0} \, , \\
  \Gamma_\pm^{(1)} &=& \frac{\kappa g_1^2 + (\gamma_1+\gamma_2)\left(\epsilon_\pm^{(1)}\right)^2}{g_1^2+\Omega_c^2+\left(\epsilon_\pm^{(1)}\right)^2} \, .
\end{eqnarray}
\end{subequations}
As before, all of the $\Gamma$'s could have also been calculated from $-i\hbar\Gamma_j^{(n)} = \langle e_j^{(n)}|{\mathcal H}_{res}|e_j^{(n)}\rangle$ and $-i\hbar\Gamma_{jk}^{(n)} = \langle e_j^{(n)}|{\mathcal H}_{res}|e_k^{(n)}\rangle$, but again, the outlined procedure offers a deeper physical insight.

\subsection{Quantum Jumps}
\label{sec:qjumps}

The effective Hamiltonian~(\ref{eq:heff}) describes the time evolution of the quantum system between successive jumps. The effect of quantum jumps is not included, and the proper way to include these is the subject of the quantum trajectories approach~\cite{Carmichael93B}. Here, we briefly describe the transformation of collapses into the dressed state basis.

Quantum jumps are included in the master equation for the time evolution of the density matrix $\rho$ via terms of the form $C_j\rho C_j^\dagger$, where $C_j$ denotes a collapse operator from the set~(\ref{eq:collapse}). Each of the collapse operators can then be expressed in terms of polariton operators, and a new set of collapse operators $S_{ij}^{(n)} = \sqrt{\Gamma_j^{(n)}} p^{(n)}_{ij}$ can be obtained. Note that the effective master equation resulting from the polariton expansion will contain cross terms in the collapse operators, giving damping terms of the form $\Gamma_{jk}^{(n)}\left( 2S_{ij}^{(n)}\rho S_{ik}^{(n)\dagger} - S_{ik}^{(n)\dagger}S_{ij}^{(n)}\rho - \rho S_{ik}^{(n)\dagger}S_{ij}^{(n)}\right)$. These cross terms have a very important role in modifying the emission rate from the polariton states to the reservoir. Such terms have been studied and well understood for the case of the modification of spontaneous emission in multilevel atoms~\cite{Agarwal74,Cardimona82}.

\section{Dynamic Stark Splitting}
\label{sec:dynstark}

It was proven earlier~\cite{Rebic99,Werner99} that the EIT-Kerr system can behave as an effective two-level system, with the states $|e_0^{(0)}\rangle$ and $|e_0^{(1)}\rangle$ being its ground and excited states. The two-level approximation is best for large values of effective dipole coupling, i.e. $(g_1/\Omega_c)^2 \gg 1$ and $g_2 \gg \kappa$. The two states of the effective model are coupled by the external field, with the Rabi frequency of the coupling, $\Omega_0^{(0,1)}$, given by Eq.~(\ref{eq:omega0}), and the decay rate of the excited state, $\Gamma_0^{(1)}$, given by Eq.~(\ref{eq:gamma0}). It is therefore expected that this system exhibits a dynamic Stark splitting, characteristic of every driven two-state system. We now explore this effect in more detail, using the polariton model.

\begin{figure}[!t]
   \includegraphics[scale=0.7]{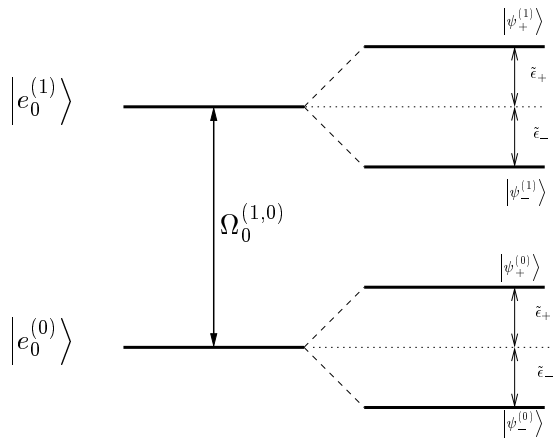}
   \caption{Graphical representation of the dynamic Stark splitting of the effective two-level system.}
   \label{fig:stark}
\end{figure}

Recall that the effective Hamiltonian~(\ref{eq:heff}) is non-Hermitian. We truncate the expansion over manifolds of the polariton Hamiltonian after the first manifold. Moreover, we concentrate on the ground and excited state of the effective two level system and write its polariton Hamiltonian in the reduced form
\begin{equation}
  \label{eq:htwolevel}
  {\mathcal H}_{red} = i\hbar\Omega_0^{(1,0)} \left( p_0^{(1)} - p_0^{(1)\dagger} \right) -i\hbar\Gamma_0^{(1)} p_0^{(1)\dagger} p_0^{(1)} \, .
\end{equation}
The eigenvalues of such an effective Hamiltonian are complex:
\begin{eqnarray}
  \label{eq:eigred}
  \varepsilon_\pm &=& \tilde{\epsilon}_\pm - i\tilde{\Gamma}_\pm \nonumber \\
  &=& i\frac{\Gamma_0^{(1)}}{2} \pm \sqrt{\Omega_0^{(1,0)^2} - \left(\Gamma_0^{(1)}/2 \right)^2} \, .
\end{eqnarray}
The real parts of these eigenvalues represent the energies of the dressed states, while the imaginary parts represent their associated decay rates. We identify two operating regimes, depending on the size of $\Omega_0^{(1,0)}$, i.e. the size of ${\mathcal E}_p$:
\begin{itemize}
\item \underline{Regime 1}: $\Omega_0^{(1,0)} < \Gamma_0^{(1)}/2$ \\
  \begin{subequations}
    \label{eq:regimes}
  \begin{eqnarray}
  \tilde{\epsilon}_\pm &=& 0 \, , \nonumber \\
  \tilde{\Gamma}_\pm &=& \frac{\Gamma_0^{(1)}}{2} \pm \sqrt{\left(\Gamma_0^{(1,0)}/2 \right)^2-\Omega_0^{(1,0)^2}}\, . \label{eq:wkdrivdec}
  \end{eqnarray}
\item \underline{Regime 2}: $\Omega_0^{(1,0)} > \Gamma_0^{(1)}/2$ \\
  \begin{eqnarray}
  \tilde{\epsilon}_\pm &=& \pm \sqrt{\Omega_0^{(1,0)^2}-\left(\Gamma_0^{(1)}/2 \right)^2} \label{eq:epm} \, , \nonumber \\
  \tilde{\Gamma}_\pm &=& \frac{\Gamma_0^{(1)}}{2}\, .
  \end{eqnarray}
  \end{subequations}
\end{itemize}
The eigenstates corresponding to the Stark-split states are
\begin{eqnarray}
  \label{eq:rabistates}
  |\psi_\pm^{(0,1)} \rangle &=& \frac{1}{\sqrt{2}} \, \bigl( |e_0^{(0)} \rangle \pm |e_0^{(1)} \rangle \bigr) \nonumber \\
  &=& \frac{1}{\sqrt{2}} \, \biggl( |0,1 \rangle \pm \frac{|1,1\rangle + g_1/\Omega_c \, |0,3 \rangle}{\sqrt{1+(g_1/\Omega_c)^2}} \biggr) \, .
\end{eqnarray}
The transition between the two regimes happens at $\Omega_0^{(1,0)} = \Gamma_0^{(1)}/2$, or, in terms of the original parameters, at
\begin{equation}
  \label{eq:regtrans}
  {\mathcal E}_p = \frac{\kappa/2}{\sqrt{1+\bigl(g_1/\Omega_c \bigr)^2}} \, .
\end{equation}
Note that there is no contribution of the atomic decay rates in this simple model. This absence occurs because the excited state in the effective model is a dark state with respect to atomic spontaneous emission. The dynamic Stark splitting effect is shown schematically in Fig.~\ref{fig:stark}.

Another effective two-level system was predicted by Tian and Carmichael~\cite{Tian92}, who studied the case of a two-level atom in the cavity, driven on the lower Rabi resonance. They also predicted Stark splitting~\cite{Carmichael94} comparable to that presented in this paper.

For further comparison, the analysis of a two-level atom with spontaneous emission rate $\gamma$, coupled with strength $g$ to the vacuum cavity mode, yields results identical to those of Eqs.~(\ref{eq:regimes}), with the replacements $\Omega_0^{(1,0)} \rightarrow g$ and $\Gamma_0^{(1)} \rightarrow \gamma+\kappa$. Thus, the dynamic Stark splitting found in the EIT-Kerr system can be thought of as the exact counterpart to the vacuum Rabi splitting characteristic for a two-level atom coupled to the cavity mode. The distinction of the EIT-Kerr system is in the fact that the parameters $\Omega_0^{(1,0)}$ and $\Gamma_0^{(1)}$ can be adjusted by a simple adjustment of the coupling laser Rabi frequency $\Omega_c$.

\begin{figure}[!t]
   \includegraphics[scale=0.55]{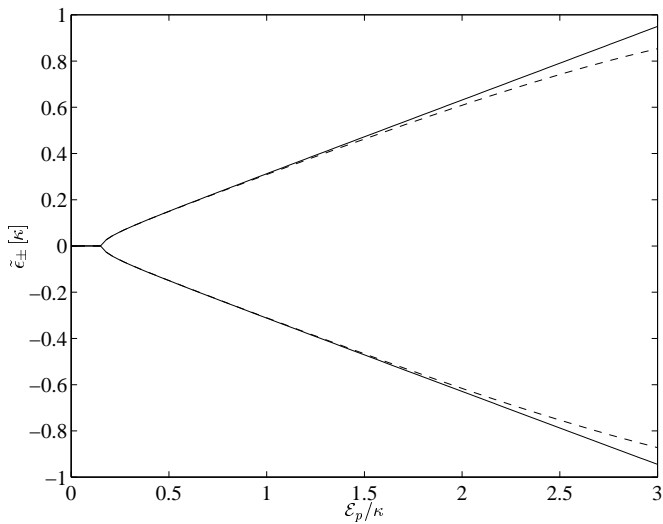}
   \caption{Numerical (dashed lines) eigenenergies for the full system (cavity mode subspace truncated at 40) compared with the anaytical solution of Eq.~(\ref{eq:eigred}) (solid lines). The parameters are $g_j/\kappa = 6$, $\gamma_j/\kappa = 0.1$, $\Omega_c/\kappa = 2$.}
   \label{fig:rabi}
\end{figure}

How well can these results describe the full EIT-Kerr system, including damping terms, as described by Hamiltonian~(\ref{eq:heff})? Fig.~\ref{fig:rabi} compares the two solutions. There is very good agreement between numerical solution and analytical approximation, which breaks down only for large values of ${\mathcal E}_p$, where truncation after the first manifold is not justified any more, since the contribution of states from higher manifolds cannot be ignored.

\begin{figure}[!t]
   \includegraphics[scale=1]{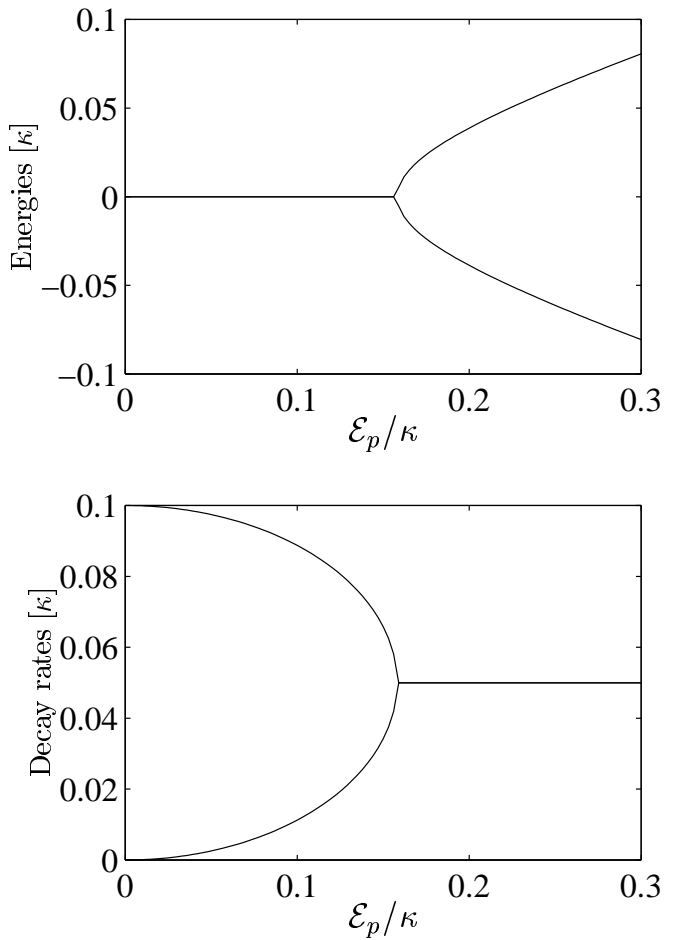}
   \caption{Eigenenergies and their associated decay rates of the Stark split states in the weak excitation regime for the same set of parameters as in Fig.~\ref{fig:rabi}.}
   \label{fig:weakex}
\end{figure}

One additional feature can be seen by looking at the eigenenergies in Figure~\ref{fig:rabi}. Notice that the Stark splitting of eigenenergies does not start at ${\mathcal E}_p = 0^+$ but at some small, finite value of ${\mathcal E}_p$. This behaviour and the related behaviour of the decay rates for weak excitation is shown in Fig.~\ref{fig:weakex}. In the limit ${\mathcal E}_p = 0$, the ground and excited states of the effective two level system are uncoupled, so the decay rates separate accordingly to the decay rates of ground state (which is zero) and the excited state $\Gamma_0^{(1)}$. Increasing the driving strength mixes these two states so that their decay rates become approximately equal. Once ${\mathcal E}_p$ exceeds the value given by Eq.~(\ref{eq:regtrans}), the energy levels shift in opposite directions, giving rise to the Stark splitting. For the parameters of Fig.~\ref{fig:weakex}, this happens at ${\mathcal E}_p \cong 0.16\kappa$.

In the weak driving regime (Regime 1 above), ${\mathcal E}_p < (\kappa/2)/\sqrt{1+\bigl(g_1/\Omega_c \bigr)^2}$, and the EIT-Kerr system truly behaves as an effective two-level system due to the absence of normal mode splitting. This is also the ideal photon blockade regime, and will be called a {\em weak driving regime}. The case ${\mathcal E}_p > (\kappa/2)/\sqrt{1+\bigl(g_1/\Omega_c \bigr)^2}$ (Regime 2) then includes the intermediate and strong driving regimes, a study of which will be published elsewhere. Once again, we emphasize the similarity of the effective two-level system coupled to the driving field and the two-level atom coupled to vacuum cavity mode. There is an equivalence in the behaviour of the Rabi-split states in the latter~\cite{Turchette95b,Kimble94} to the behaviour of the Stark-split states in the former.

The analysis of the dynamics of the Stark splitting in the dressed state basis offers a simple example of the convenience of the polariton approach. Once the Hamiltonian is expressed in terms of the polaritons and the reduced effective Hamiltonian is identified, the subsequent analysis is considerably simplified.

\section{Fluorescence Spectrum}
\label{sec:fluorescence}

The prediction of dynamic Stark splitting in Section~\ref{sec:dynstark} leads us naturally to an examination of the fluorescence spectrum of the light emitted by the `atom-cavity molecule'. We solve the master equation of the problem numerically to obtain the spectrum, and interpret the result using the insight provided by the polariton analysis.

\begin{figure}[!t]
   \includegraphics[scale=0.55]{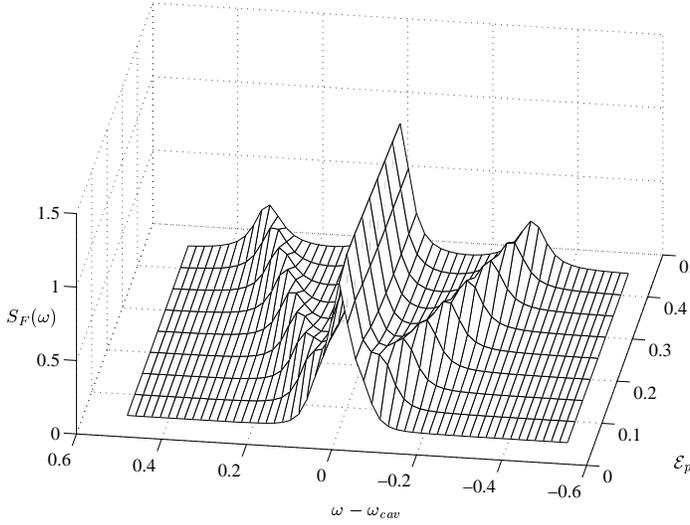}
   \caption{Emergence of Mollow triplet with the increasing driving field. Parameters are $\kappa = 0.25$, $\gamma_j = 0.1$, $g_j = 6$, $\Omega_c = 2$, $\delta = 0$ and $\Delta = 0.1$.}
   \label{fig:mollow}
\end{figure}

\begin{figure}[!t]
   \includegraphics[scale=0.55]{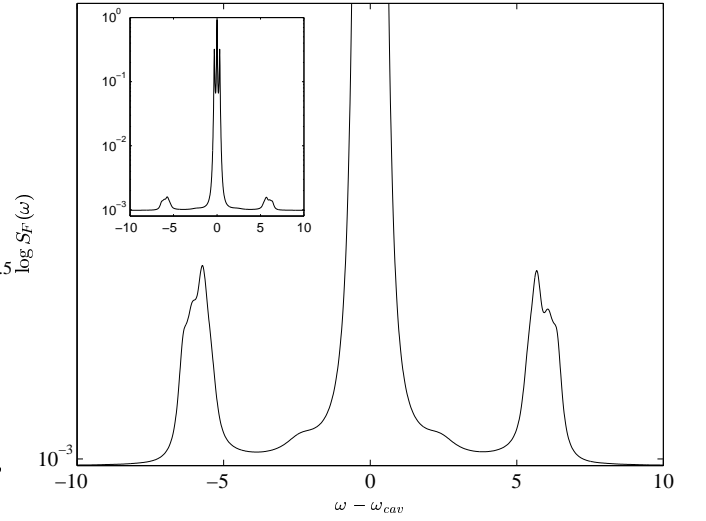}
   \caption{Semilogarithmic plot of the fluorescence spectrum for the same parameters as in Fig.~\ref{fig:mollow}, and ${\mathcal E}_p = 0.45$. Inset shows the full spectrum, while main figure shows the enlargement of the sidebands.}
   \label{fig:flspec}
\end{figure}

The master equation of the full atom/cavity system may be written in the bare form as
\begin{eqnarray}
  \dot{\rho} = -\frac{i}{\hbar} \left( {\mathcal H}_{eff}\rho - \rho{\mathcal H}_{eff}^\dagger \right) +2\sum_{i=1}^4 C_i\rho C_i^\dagger   \label{eq:master} \, ,
\end{eqnarray}
where $\rho$ is the density matrix of the system, ${\mathcal H}_{eff}$ is given by Eq.~(\ref{eq:heff}) and $C_i$ denote the four collapse operators of Eq.~(\ref{eq:collapse}). Using the quantum regression theorem~\cite{Walls94}, we solve the master equation and calculate the steady-state fluorescence spectrum,
\begin{equation}
  \label{eq:fluspec}
  S_F(\omega) = {\rm Re}{\left[ \lim_{t\rightarrow \infty}\int_0^\infty {\rm d}\tau \langle a^\dagger(t),\, a(t+\tau) \rangle e^{i\omega\tau} \right]}\, .
\end{equation}
Results, for different values of driving ${\mathcal E}_p$, are shown in Figs.~\ref{fig:mollow} and~\ref{fig:flspec}.

Fig.~\ref{fig:mollow} shows how the central peak (at the frequency of driving) splits into a Mollow triplet~\cite{Mollow69}. The particular values of parameters are given in dimensionless units, and the small value of $\kappa$ is chosen to produce narrow, highly-resolved peaks. For this set of parameters, Eq.~(\ref{eq:regtrans}) predicts a threshold value for the appearance of Mollow sidebands at ${\mathcal E}_p = 0.0395$. The central peak is the result of the two transitions $|\psi_-^{(0)}\rangle \leftrightarrow |\psi_-^{(1)}\rangle$ and $|\psi_+^{(0)}\rangle \leftrightarrow |\psi_+^{(1)}\rangle$ between the Stark doublet states in the ground and excited states (see Fig.~\ref{fig:stark}). Transitions $|\psi_-^{(0)}\rangle \leftrightarrow |\psi_+^{(1)}\rangle$ and $|\psi_+^{(0)}\rangle \leftrightarrow |\psi_-^{(1)}\rangle$ cause the appearance of the sidebands at frequencies $\omega = \omega_{cav}+(\tilde{\epsilon}_+-\tilde{\epsilon}_-)$ and $\omega = \omega_{cav}+(\tilde{\epsilon}_--\tilde{\epsilon}_+)$, respectively, where $\tilde{\epsilon}_\pm$ are given by Eq.~(\ref{eq:epm}).

The linewidths of these peaks can also be calculated. It is straightforward to write the master equation in the polariton picture as 
\begin{eqnarray}
  \dot{\rho} = -\frac{i}{\hbar} \left( {\mathcal H}_{red}\rho - \rho{\mathcal H}_{red}^\dagger \right) +2\Gamma_0^{(1)} p_0^{(1)}\rho p_0^{(1)\dagger}   \label{eq:masterpol} \, ,
\end{eqnarray}
where ${\mathcal H}_{red}$ is given by Eq.~(\ref{eq:htwolevel}). Equations for the density matrix elements in the basis spanned by Stark states $|\psi_\pm\rangle$ of Eq.~(\ref{eq:rabistates}) can be derived using standard methods to give
\begin{subequations}
  \label{eq:dmatels}
\begin{eqnarray}
  \dot{\rho}_{++} &=& -\Gamma_0^{(1)}\rho_{++} + \frac{\Gamma_0^{(1)}}{2} \, , \\
  \dot{\rho}_{+-} &=& -\left( \frac{3\Gamma_0^{(1)}}{2} - i\, 2\Omega_0^{(1,0)} \right)\rho_{+-} \nonumber \\
  &\ &- \frac{\Gamma_0^{(1)}}{2}\rho_{-+} -\Gamma_0^{(1)} \, .
\end{eqnarray}
\end{subequations}
From these equations, it is easy to read the spectral linewidths of the Mollow spectrum. The central peak will have linewidth $\Gamma_0^{(1)}$, while the sidebands have linewidth $3\Gamma_0^{(1)}/2$. This is consistent with the results for resonance fluorescence~\cite{Walls94}.

\begin{figure}[!t]
   \includegraphics[scale=0.75]{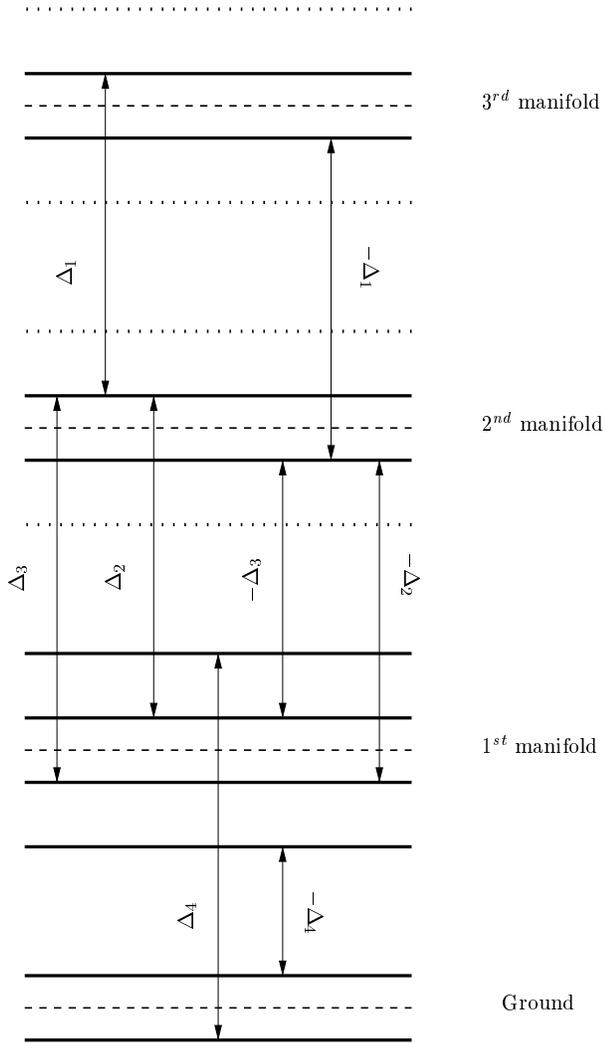}
   \caption{Schematic depiction of the relevant transitions for Fig.~\ref{fig:flspec}. Full lines denote the energy levels involved, dashed lines denote positions of the cavity resonances. $\Delta_j$'s are explained in the main text.}
   \label{fig:transflspec}
\end{figure}

Given the complex energy level structure of the atom-cavity molecule, it can be expected that transitions other than those producing the Mollow spectrum will be seen in the fluorescence spectrum. This is indeed true, and Fig.~\ref{fig:flspec} shows the additional sidebands. These peaks are relatively small $(\sim 10^{-3})$, so the associated transitions are not expected to contribute significantly to the dynamics. They do, however, cause a departure from the ideal two-level behaviour.

The transitions responsible for the sidebands are identified in Fig.~\ref{fig:transflspec}, where the relevant energy level structure is shown. Contributions of the transitions up to the third manifold states can be seen. We find peaks at the following frequencies: $\pm\Delta_1 \approx \pm 2.3$, $\pm\Delta_2 \approx \pm 5.7$, $\pm\Delta_3 \approx \pm 6.05$ and $\pm\Delta_4 \approx \pm 6.3$. The transitions corresponding to these peaks can be identified from the energy eigenvalues as $\Delta_1 = \epsilon^{(3)}_3 - \epsilon^{(2)}_3$, $-\Delta_1 = \epsilon^{(3)}_2 - \epsilon^{(2)}_2$, $\Delta_2 = \epsilon^{(2)}_3 - \tilde{\epsilon}^{(1)}_+$, $-\Delta_2 = \epsilon^{(2)}_2 - \tilde{\epsilon}^{(1)}_-$, $\Delta_3 = \epsilon^{(2)}_3 - \tilde{\epsilon}^{(1)}_-$, $-\Delta_3 = \epsilon^{(2)}_2 - \tilde{\epsilon}^{(1)}_+$, $\Delta_4 = \epsilon^{(1)}_+ - \tilde{\epsilon}^{(0)}_-$. The tiny asymmetry between the position of the positive frequency peaks and negative frequency peaks arises from the fact that the polariton states are asymmetrically detuned from the cavity resonance. The source of this asymmetry can be traced to the nonzero values of atomic detunings $\delta$ and $\Delta$ (see Fig.~\ref{fig:atom}). The linewidths of these sidebands can also be computed using the same method that produced the Eq.~(\ref{eq:dmatels}), but the results would be hard to verify, given the large degree of overlap between the adjacent peaks, seen in Fig.~\ref{fig:flspec}. We thus have a complete explanation of the fluorescence spectrum.

\section{Conclusion and Outlook}
\label{sec:conclusion}

In this paper, we have presented an exact solution to the eigenvalue problem of the Hamiltonian for a four-level atom strongly coupled to a cavity mode. The regime of strong coupling CQED presents a difficult problem for analytical calculation, as well as for the understanding of the physics involved. We have shown that consistent application of the polariton approach can offer a significant insight and even enable a relatively simple analytical treatment of the physical problem. In particular, the problem of an externally driven atom/quantum field system has been reduced to the problem of composite excitations, transitions between which are effectively driven by classical fields of Rabi frequencies $\Omega_{ij}^{(n,n-1)}$, which were calculated exactly.

The polariton approach can be interpreted as a change of basis in Hilbert space. It is obvious that such a change can in general simplify the analysis of the problem. The reason for this is that in the dressed state basis the number of degrees of freedom can be significantly reduced, compared to the treatment in terms of the bare states. For example, if the atom has $N_a$ levels (degrees of freedom), and quantum field mode can be safely truncated at some number $N_c$, the problem in the bare state basis has a dimension of at least $N_a \times N_c$. In the dressed state basis, we can identify which $N_p$ dressed states (and associated polaritons) participate in the dynamics, effectively reducing the dimension of the problem to $N_p \leq N_a \times N_c$. The treatment of dynamic Stark splitting in Section~\ref{sec:dynstark} provided a simple but extremely successful example of such reduction. In other words, from all of the (infinite number of) dimensions of Hilbert space, the polariton approach lets us pinpoint those few dimensions that are predominantly involved in the system dynamics.

Of course, this approach does not guarantee that the reduced problem will be analytically solvable. There are some general limits on solvability in the dressed state basis. For example, if the number of atomic levels $N_a > 4$, the diagonalisation of the interaction Hamiltonian is impossible {\em in principle}, except perhaps in some special cases, since $N_a$ determines the order of the polynomial of the eigenvalue problem. Also, the size of the reduced problem $N_p$ can still be impractically large. While nothing much can be done about the first problem, for the second one, we foresee ways to simplify the involved numerics. In particular, the coupled amplitudes approach and the effective master equation look promising. We are pursuing this avenue at the moment, and will publish our findings elsewhere. 

\begin{acknowledgments}
The authors would like to thank S. G. Clark for valuable discussion. This work was supported by the Marsden Fund of the Royal Society of New Zealand.
\end{acknowledgments}

\appendix

\section{Polariton Operators of the First Manifold}
\label{sec:app}

In this Appendix, polariton operators for the first manifold states are given explicitly, and their commutation relations discussed.

From expressions~(\ref{eq:e00}) --~(\ref{eq:firstcoeff}), we deduce the form of the polariton operators in terms of atomic and field operators
\begin{subequations}
  \label{eq:polaritons}
\begin{eqnarray}
  p_0^{(1)\dagger} &=& \frac{a^{\dagger}+(g_1/\Omega_c)\sigma_{31}}{\sqrt{1+(g_1/\Omega_c)^2}} \label{eq:pol} \, , \\
  p^{(1)\dagger}_\pm &=& -\frac{\bigl(g_1/\Omega_c\bigr) \, a^\dagger + i \, \bigl(\epsilon^{(1)}_\pm/\Omega_c\bigr) \, \sigma_{21} - \sigma_{31}}{\sqrt{1+\bigl(g_1/\Omega_c\bigr)^2+\bigl(\epsilon^{(1)}_\pm/\hbar\Omega _c\bigr)^2}} \label{eq:pols13} \, .
\end{eqnarray}
\end{subequations}
It is a well-known fact that the polaritons are neither bosons nor fermions. The commutation relations satisfied by the operators $p_j^{(1)}$ and $p_j^{(1)\dagger}$ are
\begin{subequations}
  \label{eq:commrel}
\begin{eqnarray}
  \bigl[ p_0^{(1)},\, p_0^{(1)\dagger} \bigr] &=& \frac{1-\bigl( g_1/\Omega_c \bigr)^2 D_{31}}{1+\bigl( g_1/\Omega_c \bigr)^2}\, , \\
  \bigl[ p_\pm^{(1)},\, p_\pm^{(1)\dagger} \bigr] &=& \frac{\bigl( g_1/\Omega_c \bigr)^2 + \bigl( \epsilon_\pm^{(1)}/\hbar\Omega_c \bigr)^2 D_{21} - D_{31}}{1+\bigl( g_1/\Omega_c \bigr)^2+\bigl( \epsilon_\pm^{(1)}/\hbar\Omega_c \bigr)^2}\, ,
\end{eqnarray}
\end{subequations}
with $D_{21} = \sigma_{22} - \sigma_{11}$, $D_{31} = \sigma_{33} - \sigma_{11}$. Therefore, strong coupling of bosons and fermions yields an excitation (or a quasiparticle) of mixed statistics.

We can define two limits in which the polaritons become dominated by the contribution of their constituents, according to the ratio $g_1/\Omega_c$. When $g_1/\Omega_c \ll 1$, polariton $p_0^{(1)}$, and its corresponding eigenstate $|e_0^{(1)}\rangle$, become dominated by their photonic contribution, while polaritons $p_\pm^{(1)}$ and corresponding eigenstates become dominated by their atomic contribution. So, different polaritons become either photon-like or atom-like. The opposite situation occurs for $g_1/\Omega_c \gg 1$. The region $g_1/\Omega_c \sim 1$ is a ``no-mans land'', where photons and atoms contribute comparably to the composition of polaritons.

\printfigures*

\begin{thebibliography}{10}
\expandafter\ifx\csname bibnamefont\endcsname\relax
  \def\bibnamefont#1{#1}\fi
\expandafter\ifx\csname bibfnamefont\endcsname\relax
  \def\bibfnamefont#1{#1}\fi
\expandafter\ifx\csname url\endcsname\relax
  \def\url#1{\texttt{#1}}\fi
\expandafter\ifx\csname urlprefix\endcsname\relax\def\urlprefix{URL }\fi
\providecommand{\bibinfo}[2]{#2}
\providecommand{\eprint}[2][]{\url{#2}}

\bibitem{Jaynes63}
\bibinfo{author}{\bibfnamefont{E.~T.} \bibnamefont{Jaynes}} \bibnamefont{and}
  \bibinfo{author}{\bibfnamefont{F.~W.} \bibnamefont{Cummings}},
  \bibinfo{journal}{Proc. IEEE} \textbf{\bibinfo{volume}{51}},
  \bibinfo{pages}{89} (\bibinfo{year}{1963}).

\bibitem{Berman94}
\bibinfo{editor}{\bibfnamefont{P.~R.} \bibnamefont{Berman}}, ed.,
  \emph{\bibinfo{title}{Cavity Quantum Electrodynamics}}, Advances in Atomic,
  Molecular and Optical Physics, Supplement 2 (\bibinfo{publisher}{Academic},
  \bibinfo{address}{New York}, \bibinfo{year}{1994}).

\bibitem{Hood98}
\bibinfo{author}{\bibfnamefont{C.~J.} \bibnamefont{Hood}},
  \bibinfo{author}{\bibfnamefont{M.~S.} \bibnamefont{Chapman}},
  \bibinfo{author}{\bibfnamefont{T.~W.} \bibnamefont{Lynn}}, \bibnamefont{and}
  \bibinfo{author}{\bibfnamefont{H.~J.} \bibnamefont{Kimble}},
  \bibinfo{journal}{Phys. Rev. Lett.} \textbf{\bibinfo{volume}{80}},
  \bibinfo{pages}{4157} (\bibinfo{year}{1998}).

\bibitem{Ye99}
\bibinfo{author}{\bibfnamefont{J.}~\bibnamefont{Ye}},
  \bibinfo{author}{\bibfnamefont{D.~W.} \bibnamefont{Vernooy}},
  \bibnamefont{and} \bibinfo{author}{\bibfnamefont{H.~J.}
  \bibnamefont{Kimble}}, \bibinfo{journal}{Phys. Rev. Lett.}
  \textbf{\bibinfo{volume}{83}}, \bibinfo{pages}{4987} (\bibinfo{year}{1999}).

\bibitem{Guthohrlein01}
\bibinfo{author}{\bibfnamefont{G.~R.} \bibnamefont{Guth\"{o}hrlein}},
  \bibinfo{author}{\bibfnamefont{M.}~\bibnamefont{Keller}},
  \bibinfo{author}{\bibfnamefont{K.}~\bibnamefont{Hayasaka}},
  \bibinfo{author}{\bibfnamefont{W.}~\bibnamefont{Lange}}, \bibnamefont{and}
  \bibinfo{author}{\bibfnamefont{H.}~\bibnamefont{Walther}},
  \bibinfo{journal}{Nature} \textbf{\bibinfo{volume}{414}}, \bibinfo{pages}{49}
  (\bibinfo{year}{2001}).

\bibitem{Hood00}
\bibinfo{author}{\bibfnamefont{C.~J.} \bibnamefont{Hood}},
  \bibinfo{author}{\bibfnamefont{T.~W.} \bibnamefont{Lynn}},
  \bibinfo{author}{\bibfnamefont{A.~C.} \bibnamefont{Doherty}},
  \bibinfo{author}{\bibfnamefont{A.~S.} \bibnamefont{Parkins}},
  \bibnamefont{and} \bibinfo{author}{\bibfnamefont{H.~J.}
  \bibnamefont{Kimble}}, \bibinfo{journal}{Science}
  \textbf{\bibinfo{volume}{287}}, \bibinfo{pages}{1447} (\bibinfo{year}{2000}).

\bibitem{Pinkse00}
\bibinfo{author}{\bibfnamefont{P.~W.~H.} \bibnamefont{Pinkse}},
  \bibinfo{author}{\bibfnamefont{T.}~\bibnamefont{Fischer}},
  \bibinfo{author}{\bibfnamefont{P.}~\bibnamefont{Maunz}}, \bibnamefont{and}
  \bibinfo{author}{\bibfnamefont{G.}~\bibnamefont{Rempe}},
  \bibinfo{journal}{Nature} \textbf{\bibinfo{volume}{404}},
  \bibinfo{pages}{365} (\bibinfo{year}{2000}).

\bibitem{Dunstan98}
\bibinfo{author}{\bibfnamefont{M.}~\bibnamefont{Dunstan}},
  \bibinfo{author}{\bibfnamefont{S.}~\bibnamefont{Rebi\'{c}}},
  \bibinfo{author}{\bibfnamefont{S.~M.} \bibnamefont{Tan}},
  \bibinfo{author}{\bibfnamefont{A.~S.} \bibnamefont{Parkins}},
  \bibinfo{author}{\bibfnamefont{M.~J.} \bibnamefont{Collett}},
  \bibnamefont{and} \bibinfo{author}{\bibfnamefont{D.~F.} \bibnamefont{Walls}},
  in \emph{\bibinfo{booktitle}{Proceedings: Quantum Communication, Computing
  and Measurement 2}}, edited by
  \bibinfo{editor}{\bibfnamefont{P.}~\bibnamefont{Kumar}},
  \bibinfo{editor}{\bibfnamefont{G.~M.} \bibnamefont{D'Ariano}},
  \bibnamefont{and} \bibinfo{editor}{\bibfnamefont{O.}~\bibnamefont{Hirota}}
  (\bibinfo{publisher}{Plenum Press}, \bibinfo{address}{New York},
  \bibinfo{year}{1998}).

\bibitem{Brune94}
\bibinfo{author}{\bibfnamefont{M.}~\bibnamefont{Brune}},
  \bibinfo{author}{\bibfnamefont{P.}~\bibnamefont{Nussenzveig}},
  \bibinfo{author}{\bibfnamefont{F.}~\bibnamefont{Schmidt-Kaler}},
  \bibinfo{author}{\bibfnamefont{F.}~\bibnamefont{Bernardot}},
  \bibinfo{author}{\bibfnamefont{A.}~\bibnamefont{Maali}},
  \bibinfo{author}{\bibfnamefont{J.~M.} \bibnamefont{Raimond}},
  \bibnamefont{and} \bibinfo{author}{\bibfnamefont{S.}~\bibnamefont{Haroche}},
  \bibinfo{journal}{Phys. Rev. Lett.} \textbf{\bibinfo{volume}{72}},
  \bibinfo{pages}{3339} (\bibinfo{year}{1994}).

\bibitem{Turchette95}
\bibinfo{author}{\bibfnamefont{Q.~A.} \bibnamefont{Turchette}},
  \bibinfo{author}{\bibfnamefont{C.~J.} \bibnamefont{Hood}},
  \bibinfo{author}{\bibfnamefont{W.}~\bibnamefont{Lange}},
  \bibinfo{author}{\bibfnamefont{H.}~\bibnamefont{Mabuchi}}, \bibnamefont{and}
  \bibinfo{author}{\bibfnamefont{H.~J.} \bibnamefont{Kimble}},
  \bibinfo{journal}{Phys. Rev. Lett.} \textbf{\bibinfo{volume}{75}},
  \bibinfo{pages}{4710} (\bibinfo{year}{1995}).

\bibitem{Harris97}
\bibinfo{author}{\bibfnamefont{S.~E.} \bibnamefont{Harris}},
  \bibinfo{journal}{Physics Today} \textbf{\bibinfo{volume}{50}},
  \bibinfo{pages}{36} (\bibinfo{year}{1997}).

\bibitem{Schmidt96}
\bibinfo{author}{\bibfnamefont{H.}~\bibnamefont{Schmidt}} \bibnamefont{and}
  \bibinfo{author}{\bibfnamefont{A.}~\bibnamefont{Imamo\u{g}lu}},
  \bibinfo{journal}{Opt. Lett.} \textbf{\bibinfo{volume}{21}},
  \bibinfo{pages}{1936} (\bibinfo{year}{1996}).

\bibitem{Imamoglu97}
\bibinfo{author}{\bibfnamefont{A.}~\bibnamefont{Imamo\u{g}lu}},
  \bibinfo{author}{\bibfnamefont{H.}~\bibnamefont{Schmidt}},
  \bibinfo{author}{\bibfnamefont{G.}~\bibnamefont{Woods}}, \bibnamefont{and}
  \bibinfo{author}{\bibfnamefont{M.}~\bibnamefont{Deutsch}},
  \bibinfo{journal}{Phys. Rev. Lett.} \textbf{\bibinfo{volume}{79}},
  \bibinfo{pages}{1467} (\bibinfo{year}{1997}).

\bibitem{Rebic99}
\bibinfo{author}{\bibfnamefont{S.}~\bibnamefont{Rebi\'{c}}},
  \bibinfo{author}{\bibfnamefont{S.~M.} \bibnamefont{Tan}},
  \bibinfo{author}{\bibfnamefont{A.~S.} \bibnamefont{Parkins}},
  \bibnamefont{and} \bibinfo{author}{\bibfnamefont{D.~F.} \bibnamefont{Walls}},
  \bibinfo{journal}{J. Opt. B: Quant. Semiclass. Opt.}
  \textbf{\bibinfo{volume}{1}}, \bibinfo{pages}{490} (\bibinfo{year}{1999}).

\bibitem{Werner99}
\bibinfo{author}{\bibfnamefont{M.~J.} \bibnamefont{Werner}} \bibnamefont{and}
  \bibinfo{author}{\bibfnamefont{A.}~\bibnamefont{Imamo\u{g}lu}},
  \bibinfo{journal}{Phys. Rev. A} \textbf{\bibinfo{volume}{61}},
  \bibinfo{pages}{011801(R)} (\bibinfo{year}{1999}).

\bibitem{Greentree00}
\bibinfo{author}{\bibfnamefont{A.~D.} \bibnamefont{Greentree}},
  \bibinfo{author}{\bibfnamefont{J.~A.} \bibnamefont{Vaccaro}},
  \bibinfo{author}{\bibfnamefont{S.~R.} \bibnamefont{de~Echaniz}},
  \bibinfo{author}{\bibfnamefont{A.~V.} \bibnamefont{Durant}},
  \bibnamefont{and} \bibinfo{author}{\bibfnamefont{J.~P.}
  \bibnamefont{Marangos}}, \bibinfo{journal}{J. Opt. B: Quantum Semiclass.
  Opt.} \textbf{\bibinfo{volume}{2}}, \bibinfo{pages}{252}
  (\bibinfo{year}{2000}).

\bibitem{Fleischhauer00}
\bibinfo{author}{\bibfnamefont{M.}~\bibnamefont{Fleischhauer}},
  \bibinfo{author}{\bibfnamefont{S.~F.} \bibnamefont{Yelin}}, \bibnamefont{and}
  \bibinfo{author}{\bibfnamefont{M.~D.} \bibnamefont{Lukin}},
  \bibinfo{journal}{Opt. Comm.} \textbf{\bibinfo{volume}{179}},
  \bibinfo{pages}{395} (\bibinfo{year}{2000}).

\bibitem{Fleischhauer00b}
\bibinfo{author}{\bibfnamefont{M.}~\bibnamefont{Fleischhauer}}
  \bibnamefont{and} \bibinfo{author}{\bibfnamefont{M.~D.} \bibnamefont{Lukin}},
  \bibinfo{journal}{Phys. Rev. Lett.} \textbf{\bibinfo{volume}{84}},
  \bibinfo{pages}{5094} (\bibinfo{year}{2000}).

\bibitem{Fleischhauer01}
\bibinfo{author}{\bibfnamefont{M.}~\bibnamefont{Fleischhauer}}
  \bibnamefont{and} \bibinfo{author}{\bibfnamefont{M.~D.} \bibnamefont{Lukin}},
    \bibinfo{howpublished}{preprint, {\tt quant-ph/0106066}}.

\bibitem{Juzeliunas01}
\bibinfo{author}{\bibfnamefont{G.}~\bibnamefont{Juzeliunas}} \bibnamefont{and}
  \bibinfo{author}{\bibfnamefont{H.~J.} \bibnamefont{Carmichael}},
  \bibinfo{howpublished}{preprint, {\tt quant-ph/0107053}}.

\bibitem{Carmichael93B}
\bibinfo{author}{\bibfnamefont{H.~J.} \bibnamefont{Carmichael}},
  \emph{\bibinfo{title}{An Open System Approach to Quantum Optics}}, vol.
  \bibinfo{volume}{M18} of \emph{\bibinfo{series}{Lecture Notes in Physics}}
  (\bibinfo{publisher}{Springer}, \bibinfo{address}{Berlin},
  \bibinfo{year}{1993}).

\bibitem{Cohen77}
\bibinfo{author}{\bibfnamefont{C.}~\bibnamefont{Cohen-Tannoudji}}
  \bibnamefont{and} \bibinfo{author}{\bibfnamefont{S.}~\bibnamefont{Reynaud}},
  \bibinfo{journal}{J. Phys. B: At. Mol. Phys.} \textbf{\bibinfo{volume}{10}},
  \bibinfo{pages}{345} (\bibinfo{year}{1977}).

\bibitem{Alsing92}
\bibinfo{author}{\bibfnamefont{P.}~\bibnamefont{Alsing}},
  \bibinfo{author}{\bibfnamefont{D.-S.} \bibnamefont{Guo}}, \bibnamefont{and}
  \bibinfo{author}{\bibfnamefont{H.~J.} \bibnamefont{Carmichael}},
  \bibinfo{journal}{Phys. Rev. A} \textbf{\bibinfo{volume}{45}},
  \bibinfo{pages}{5135} (\bibinfo{year}{1992}).

\bibitem{Harris89}
\bibinfo{author}{\bibfnamefont{S.~E.} \bibnamefont{Harris}},
  \bibinfo{journal}{Phys. Rev. Lett.} \textbf{\bibinfo{volume}{62}},
  \bibinfo{pages}{1033} (\bibinfo{year}{1989}).

\bibitem{Imamoglu89a}
\bibinfo{author}{\bibfnamefont{A.}~\bibnamefont{Imamo\u{g}lu}},
  \bibinfo{journal}{Phys. Rev. A} \textbf{\bibinfo{volume}{40}},
  \bibinfo{pages}{2835} (\bibinfo{year}{1989}).

\bibitem{Li95}
\bibinfo{author}{\bibfnamefont{Y.-Q.} \bibnamefont{Li}} \bibnamefont{and}
  \bibinfo{author}{\bibfnamefont{M.}~\bibnamefont{Xiao}},
  \bibinfo{journal}{Phys. Rev. A} \textbf{\bibinfo{volume}{51}},
  \bibinfo{pages}{4959} (\bibinfo{year}{1995}).

\bibitem{Zhou97}
\bibinfo{author}{\bibfnamefont{P.}~\bibnamefont{Zhou}} \bibnamefont{and}
  \bibinfo{author}{\bibfnamefont{S.}~\bibnamefont{Swain}},
  \bibinfo{journal}{Phys. Rev. A} \textbf{\bibinfo{volume}{56}},
  \bibinfo{pages}{3011} (\bibinfo{year}{1997}).

\bibitem{Akram01}
\bibinfo{author}{\bibfnamefont{U.}~\bibnamefont{Akram}},
  \bibinfo{author}{\bibfnamefont{Z.}~\bibnamefont{Ficek}}, \bibnamefont{and}
  \bibinfo{author}{\bibfnamefont{S.}~\bibnamefont{Swain}}, \bibinfo{journal}{J.
  Mod. Opt.} \textbf{\bibinfo{volume}{48}}, \bibinfo{pages}{1059}
  (\bibinfo{year}{2001}).

\bibitem{Agarwal74}
\bibinfo{author}{\bibfnamefont{G.~S.} \bibnamefont{Agarwal}}, in
  \emph{\bibinfo{booktitle}{Springer Tracts in Modern Physics}}, edited by
  \bibinfo{editor}{\bibfnamefont{G.}~\bibnamefont{Hohler}}
  (\bibinfo{publisher}{Springer--Verlag}, \bibinfo{address}{Berlin},
  \bibinfo{year}{1974}), vol.~\bibinfo{volume}{70}.

\bibitem{Cardimona82}
\bibinfo{author}{\bibfnamefont{D.~A.} \bibnamefont{Cardimona}},
  \bibinfo{author}{\bibfnamefont{M.~G.} \bibnamefont{Raymer}},
  \bibnamefont{and} \bibinfo{author}{\bibfnamefont{C.~R.}
  \bibnamefont{Stroud}}, \bibinfo{journal}{J. Phys. B: At. Mol. Phys.}
  \textbf{\bibinfo{volume}{15}}, \bibinfo{pages}{55} (\bibinfo{year}{1982}).

\bibitem{Tian92}
\bibinfo{author}{\bibfnamefont{L.}~\bibnamefont{Tian}} \bibnamefont{and}
  \bibinfo{author}{\bibfnamefont{H. J.}~\bibnamefont{Carmichael}},
  \bibinfo{journal}{Phys. Rev. A} \textbf{\bibinfo{volume}{46}},
  \bibinfo{pages}{R6801} (\bibinfo{year}{1992}).

\bibitem{Carmichael94}
\bibinfo{author}{\bibfnamefont{H.~J.} \bibnamefont{Carmichael}},
  \bibinfo{author}{\bibfnamefont{L.}~\bibnamefont{Tian}},
  \bibinfo{author}{\bibfnamefont{W.}~\bibnamefont{Ren}}, \bibnamefont{and}
  \bibinfo{author}{\bibfnamefont{P.}~\bibnamefont{Alsing}}, in
  \bibinfo{editor}{\bibnamefont{\bibinfo{editor}{Berman}}}  \cite{Berman94}.

\bibitem{Turchette95b}
\bibinfo{author}{\bibfnamefont{Q.~A.} \bibnamefont{Turchette}},
  \bibinfo{author}{\bibfnamefont{R.~J.} \bibnamefont{Thompson}},
  \bibnamefont{and} \bibinfo{author}{\bibfnamefont{H.~J.}
  \bibnamefont{Kimble}}, \bibinfo{journal}{Appl. Phys. B}
  \textbf{\bibinfo{volume}{60}}, \bibinfo{pages}{S1} (\bibinfo{year}{1995}).

\bibitem{Kimble94}
\bibinfo{author}{\bibfnamefont{H.~J.} \bibnamefont{Kimble}}, in
  \bibinfo{editor}{\bibnamefont{\bibinfo{editor}{Berman}}}  \cite{Berman94}.

\bibitem{Walls94}
\bibinfo{author}{\bibfnamefont{D.~F.} \bibnamefont{Walls}} \bibnamefont{and}
  \bibinfo{author}{\bibfnamefont{G.~J.} \bibnamefont{Milburn}},
  \emph{\bibinfo{title}{Quantum Optics}} (\bibinfo{publisher}{Springer},
  \bibinfo{address}{Berlin}, \bibinfo{year}{1994}).

\bibitem{Mollow69}
\bibinfo{author}{\bibfnamefont{B.~R.} \bibnamefont{Mollow}},
  \bibinfo{journal}{Phys. Rev.} \textbf{\bibinfo{volume}{188}},
  \bibinfo{pages}{1969} (\bibinfo{year}{1969}).

\end{thebibliography}
\end{document}